# A Hypothetical Investigation into the Realm of the Microscopic and Macroscopic universes beyond the Standard Model

**Balungi Francis**

**Faculty of Technology, Makerere University, Kampala Uganda, Box 7062**

**Abstract**

In an attempt to merge the microscopic with the macroscopic worlds, we present a brief study about a force which depends on the Planck force and on the coupling constant that in turn depends on the size of a particle in a particular direction. Such a force is given as, $s\delta^{n-1}$ where s, is the Planck force, $\delta$-coupling constant and n is an integer, which determines the size of a particle or an object. We then apply this force to black body radiations from which we deduce Hawking Radiations, Stefan's radiation law and the level at which the profound theories of physics are consistent. In conclusion, it true that the world exists in both scopic situations, in away that, the higher the value of n the smaller the size of the particles we are studying and vice versa is true.

## 1. Introduction

With the introduction of string theory, most of the physicists around the world have agreed that such a theory clearly describes all the fundamental interactions. Investigations are into determining whether the extra dimensions given by the theory, truly describe the microscopic world. This will require a comparison between the macroscopic with the microscopic objects. This can be worked out only if we create approximations which are clear and simple.

Such approximations will require us to study the interactions of elementary particles in terms of the strength of the interactions or forces in different space dimensions. In this attempt the force, B, is seen to depend on the coupling constant, δ, which in turn depends on the dimensions of space (n-1), where, n, is an a number that determines the size of a particle in a particular direction ( i.e. dimension of space). In other words the force does not depend on the mass of the particles or on the distance between them, but on the ability to act depending on the particles exchanged between the elementary particles and the dimensions of space in which such particles can be fully determined. Mathematically this can be shown as,

$$B_n = s\delta^{n-1} \tag{1}$$

Where s, is a constant force for all interactions, it can also be defined as a force that defines the curvature of space in which a particle moves. This curvature can then be seen to depend on the distribution of matter and energy in space. Therefore matter can now best be understood in terms of a force given by Eq1, which depends on another constant force that determines the curved nature of space time in which a particle moves and the position or location of such a particle in space defined, by n.

From Eq1, it can be clearly seen, that, s, is to macroscopic as, $\delta^{n-1}$ is to the microscopic. With the two combined, we create a theory that deals with a point particle in space time. In other words, we create a universe that is zero in size and that extends to higher space dimensions.

## 2. Hawking radiations and Everything

### 2.1. Power of the radiations

Eq1 can be applied to Black – Body radiations and to space radiations. If, in black body radiations the Energy of the photon is related to its frequency a, of the light associated with it, then also it is true, that the power, P, or the rate of change of energy of the photon is related to the square of this frequency, $a^2$, This frequency, increases with $B_n$ but falls off as the momentum of the Photon γ. Then $a = \frac{B_n}{\gamma}$, this means, that, the higher the momentum of a particle, the smaller the frequency of the radiations and vice versa. Then the power of the radiations is;

$$P_n = \hbar \frac{B_n^2}{\gamma^2} = \hbar \frac{s^2 \delta^{2(n-1)}}{\gamma^2} \qquad (2)$$

Where ℏ is the reduced Planck constant,

### 2.2. Time-scopic

Since the time t, taken for a body to emit the radiations increases with γ, but falls with $B_n$, then for all radiations the time is given by

$$t_n = \frac{\gamma}{B_n} = \frac{\gamma}{s} \delta^{1-n} \qquad (3)$$

### 2.3. Energy-scopic

But the energy of the radiations is power multiplied by time, then energy is given as,

$$w_n = P_n t_n = \frac{\hbar s}{\gamma} \delta^{n-1} \qquad (4)$$

### 2.4. Temperature-scopic

Let us assume that the energy of the radiations Eq4 is equal to the random kinetic energy of the particle at a given temperature T. then this temperature is the temperature of the body emitting the radiations and is given by

$$T_n = \frac{\hbar s}{k\gamma} \delta^{n-1} \tag{5}$$

Where, k, is the gas constant per mole
+

### 2.5. Entropy -scopic

If the particle exhibits wave properties, then at a given wavelength λ, the momentum $\gamma = \frac{\hbar}{\lambda}$ and then the temperature given by Eq4 become;

$$T_n' = \frac{\lambda s}{k} \delta^{n-1}$$

But for γ= mc, that is, for a particle of mass m, moving at a speed of light c, the time given by Eq3, becomes

$$t_n' = \frac{mc}{s} \delta^{1-n}$$

And the new energy then is obtained as, $w_n' = P_n t_n' = \frac{\hbar m c s}{\gamma^2} \delta^{n-1}$, this energy equation is a result of combining quantum mechanics with relativistic motions which gives relativistic - quantum mechanics. Then entropy **S** is the ratio of this energy to the temperature

$$S = \frac{w_n'}{T_n'} = \frac{\hbar k m c}{\gamma^2 \lambda} \tag{6}$$

Therefore, the Entropy of a system increases with the momentum of a relativistic particle whose mass is known and falls off with the square of the momentum of the massless particle.

For wavelengths, λ, approaching the Schwarzschild radius (i.e. $\lambda = \frac{Gm}{c^2}$ where G is the universal gravitational constant), and for a momentum γ, falling off with the radius, R, of the surface of the body emitting the radiations $\gamma = \frac{\hbar}{R}$, then the Entropy S, is given by

$$S = \frac{kAc^3}{\hbar G} = \frac{Ak}{l_p^2} \tag{7}$$

Where, A is the surface area of the body $R^2$ and $l_p$ is Planck length $\sqrt{\frac{\hbar G}{c^3}}$.

From the derivations given above, this is the Entropy of the Black Hole, which was predicted by Hawking.

### 3. On Stefan's Radiation Law

From Eq2, if we set n =3 , for a sphere emitting the radiations and at whose surface the particles have a momentum $\gamma = \frac{\hbar}{R}$, so that the power is,

$$P_3 = \frac{R^2 s^2}{\hbar} \delta^4$$

But then the energy of the radiations can only be studied in 2 dimensions (i.e. n=2), from Eq4, energy at 2 dimensions is;

$$w_2 = \frac{\hbar s}{\gamma} \delta$$

From which the coupling constant $\delta = \frac{w_2 \gamma}{\hbar s}$, so that, when we substitute it in $P_3$ we obtain, the power of the radiations as;

$$P_3 = \frac{R^2 s^2}{\hbar} \left[\frac{w_2 \gamma}{\hbar s}\right]^4$$

$$P_3 = \frac{R^2 w_2^4}{\hbar^5 s^2} \gamma^4$$

But for $\gamma$=mc, and m is the Planck mass $\sqrt{\frac{\hbar c}{G}}$ and $w_2$ is the random kinetic energy of the particles ( kT), and $s = \frac{8\sqrt{15}c^4}{4\pi G}$, then

$$P_3 = \frac{16\pi^2 R^2 k^4 T^4 G^2}{960\hbar^5 c^8}\left[c\sqrt{\frac{\hbar c}{G}}\right]^4$$

On arranging, we obtain

$$P_3 = R^2\left(\frac{\pi^2 k^4}{60\hbar^3 c^2}\right)T^4$$

Where a quantity $\frac{\pi^2 k^4}{60\hbar^3 c^2}$ is the Stefan- Boltzmann constant

It is therefore true that quantum effects allow black holes to emit exact black body radiations

## 4. Space Dimensions at which our Theories are Consistent

Using the above Equations for power, time, temperature and Energy, We can set dimensions at which all the theories of physics are consistent. For example, we can set $s = \frac{c^4}{G}$ as the Planck force, $\delta = \frac{Gm^2}{\hbar c}$ as the gravitational coupling constant and γ=mc as the momentum of a relativistic particle. Then

At n = 0

$$P_0 = \frac{\hbar^3 c^8}{G^4 m^6}$$

$$t_0 = \frac{G^2 m^3}{\hbar c^4}$$

$$T_0 = \frac{\hbar^2 c^4}{G^2 m^3 k}$$

$$w_0 = \frac{\hbar^2 c^4}{G^2 m^3}$$

At n=1

$$P_1 = \frac{\hbar c^6}{G^2 m^2}$$

$$t_1 = \frac{Gm}{c^3}$$

$$T_1 = \frac{\hbar c^3}{Gmk}$$

$$w_1 = \frac{\hbar c^3}{Gm}$$

At n=2

$$P_2 = \frac{m^2 c^4}{\hbar}$$

$$t_2 = \frac{\hbar}{mc^2}$$

$$T_2 = \frac{mc^2}{k}$$

$$w_2 = mc^2$$

At n= $\frac{3}{2}$

$$P_{3/2} = \frac{c^5}{G}$$

$$t_{3/2} = \sqrt{\frac{G\hbar}{c^5}}$$

$$T_{3/2} = \sqrt{\frac{\hbar c^5}{k^2 G}}$$

$$w_{3/2} = \sqrt{\frac{\hbar c^5}{G}}$$

It can be Cleary seen from the above Equations that $t_0$, $P_1$ and $T_1$ are the predictions of the Hawking Theory of Black Holes for macroscopic universes. Therefore at n=0, we have a theory capable of explaining Quasars as observed from the radiations obtained by exchanging particles called the gluons between the elementary particles, while at n=1, we obtain a theory capable of explaining Black Holes as observed from the radiations generated by exchanging particles called Gravitons. But at n=2, we have a theory capable of explaining phenomena with relativistic quantum effects, a theory of the quantum field. At n=2, there is a constant interchange of photons between the elementary particles and $w_2$ is the Einstein mass-energy relation.

Then at n=$\frac{3}{2}$, we have a theory capable of merging gravity with the quantum field theory. This occurs at the Planck length scale, in other words as seen from the above equations it can be logically formulated, that 3/2 lies between 1 and 2, or $1 \leq \frac{3}{2} \leq 2$. We can therefore say that "at n=o, we are dealing with a universe that is zero in size and from n= 1 up to infinity we are dealing with an expanding Universe."

It can be clearly theorized that from n=o to n=2, we are dealing with the macroscopic universe and from n=2 to higher values of n we are dealing with the microscopic universe. It is therefore true that the world exists in both scopic situations, in away that, the higher the value of n the smaller the size of the particles we are studying and vice versa is true.

## 5. Conclusion

This paper set out to understand and attempt on the unification of the microscopic effects with macroscopic effects in a self -consistent manner. It was hypothesized that through the reformulation of classical physics via the introduction of the interaction part (i.e. $\delta^{n-1}$ ) and the kinetic part (i.e. s ), the fundamental force equation for all interactions is formulated. The results obtained using the n values support the quantum theory of gravity. The research questions that were raised at the beginning have been answered. For example, one of the verifiable prediction of this paper are the properties of a black hole on the macroscopic universe (n=0, 1.). It is obviously defined that, the microscopic worlds exist at higher values of n while the macroscopic worlds exist at smaller values. Based on the results and equations from this study, it is evident that a grand unified theory of physics is possible at n=3/2, the Planck scale. By using n values in a given range one is able to study different interactions or forces. This paper recommends that the mathematical formulation of the other theories (i.e. string theory) which predict the properties of matter as represented here should be modified and their parameters reduced to make more verifiable predictions about the nature of matter at a cosmic scale.